\documentclass{Interspeech}

\interspeechcameraready

\title{Speaker-Distinguishable CTC:\\ Learning Speaker Distinction Using CTC for Multi-Talker Speech Recognition}

\author[affiliation={1}]{Asahi}{Sakuma}
\author[affiliation={1}]{Hiroaki}{Sato}
\author[affiliation={1}]{Ryuga}{Sugano}
\author[affiliation={1}]{Tadashi}{Kumano}
\author[affiliation={1}]{Yoshihiko}{Kawai}
\author[affiliation={2}]{\\Tetsuji}{Ogawa}

\affiliation{}{NHK Science and Technology Research Laboratories}{Japan}
\affiliation{}{Waseda University}{Japan}
\email{sakuma.a-fc@nhk.or.jp}
\keywords{Connectionist temporal classification, Speaker-Distinguishable CTC, serialized output training, multi-talker speech recognition}

\usepackage{comment}
\usepackage[subrefformat=parens]{subcaption}
\usepackage{multirow} 
\usepackage{arydshln}
\newcommand{\bhline}[1]{\noalign{\hrule height #1}}
\usepackage{amsmath}
\usepackage{mathtools}
\usepackage{cite}

\begin{document}

\maketitle


\begin{abstract}
This paper presents a novel framework for multi-talker automatic speech recognition without the need for auxiliary information. Serialized Output Training (SOT), a widely used approach, suffers from recognition errors due to speaker assignment failures. Although incorporating auxiliary information, such as token-level timestamps, can improve recognition accuracy, extracting such information from natural conversational speech remains challenging. To address this limitation, we propose Speaker-Distinguishable CTC (SD-CTC), an extension of CTC that jointly assigns a token and its corresponding speaker label to each frame. We further integrate SD-CTC into the SOT framework, enabling the SOT model to learn speaker distinction using only overlapping speech and transcriptions. Experimental comparisons show that multi-task learning with SD-CTC and SOT reduces the error rate of the SOT model by 26\% and achieves performance comparable to state-of-the-art methods relying on auxiliary information.
\end{abstract}


\section{Introduction}
\label{sec:intro}

Multi-talker automatic speech recognition (MT-ASR) is a technique of transcribing conversations involving multiple speakers, often with overlapping speech, into separate text streams for each speaker.
This technology has been extensively studied for applications such as meeting transcription and automatic subtitle generation for talk shows~\cite{pit-asr,pit-asr-aed1, sot}.

Existing MT-ASR approaches can be broadly classified into two categories.
The first approach employs separate decoders for each speaker.
The encoder processes overlapping speech to generate speaker-specific representations, while dedicated decoders transcribe each speaker's utterances in parallel~\cite{pit-asr-aed1, mt-rnnt-aft,whisper_ts_meng}.
Although this approach is conceptually straightforward, it often struggles to preserve semantic continuity and contextual dependencies across speaker transitions, leading to errors such as redundant transcriptions of the same utterance.
Moreover, as the number of speakers increases, the system's complexity grows, reducing its scalability.

The second approach utilizes Serialized Output Training (SOT), where transcriptions from all speakers are concatenated into a single sequence with a special symbol ``$\mathtt{{<}sc{>}}$'' marking speaker boundaries~\cite{sot}.
An Attention Encoder-Decoder (AED) model~\cite{aed} is then trained to decode this serialized sequence.
SOT offers the advantage of using the same architecture as single-speaker ASR models, making it inherently scalable to a varying number of speakers while naturally capturing the inter-speaker context.

Enhancements that train the model with auxiliary information --- such as token-level timestamps~\cite{t-sot, sa-sot, ba-sot, makishima23_interspeech,overlap_encoding ,somsred} and clean audio~\cite{somsred, sot-spk-info, spk-mask, spk_info_fan} --- alongside the primary training data of conversational speech and transcripts have achieved state-of-the-art performance~\cite{sa-sot}.
However, obtaining reliable auxiliary data in real-world scenarios remains a significant challenge.

\begin{figure*}[t]
  \centering
  \includegraphics[width=.85\linewidth]{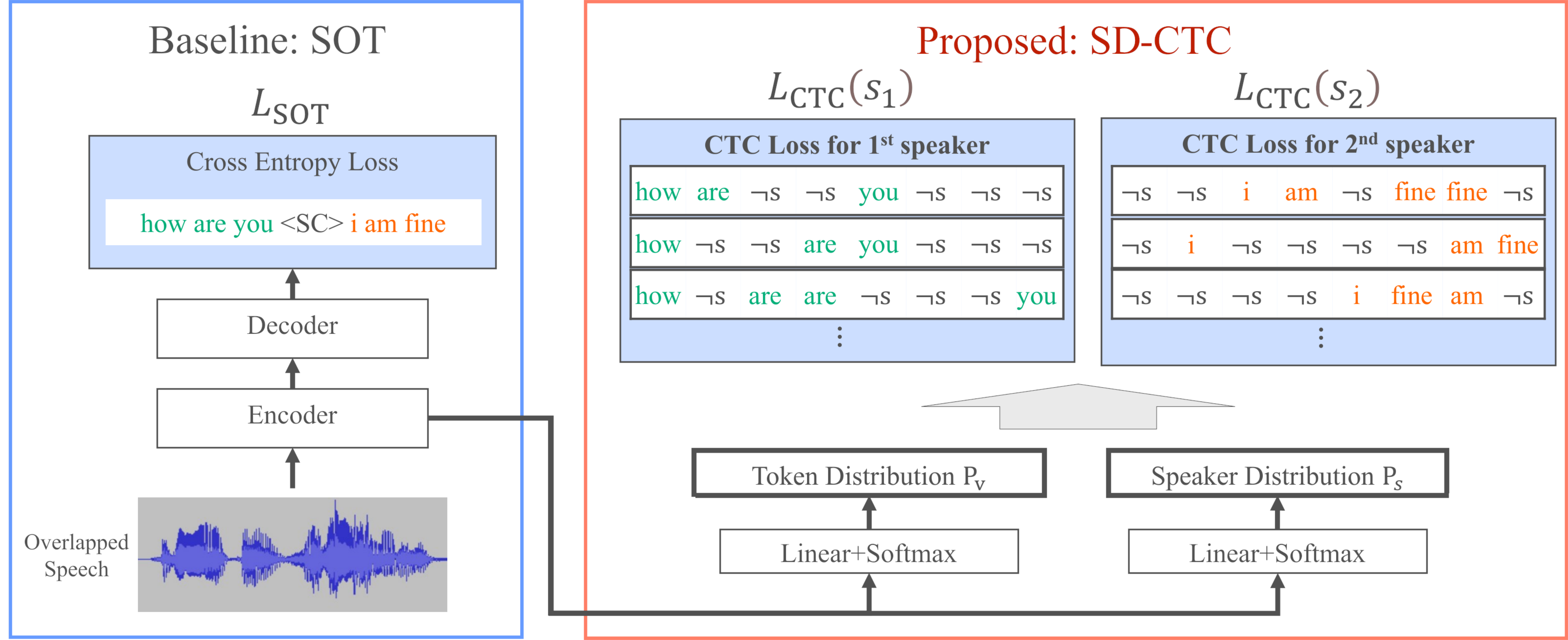}
  \vspace*{-.25cm}
  \caption{Overview of proposed method.}
  \label{overview-fig}
  \vspace{-.5cm}
\end{figure*}


Several methods have been proposed to improve SOT without auxiliary information.
They exploit multi-task learning~\cite{ctc-attention} with Connectionist Temporal Classification (CTC)~\cite{ctc} and AED models.
In one approach, multiple encoder outputs --- each corresponding to an individual speaker --- are generated, with a separate CTC loss applied to each output~\cite{gencsep}.
This enables the encoder to extract speaker-specific features, concatenate them in SOT-style order, and feed them into a single decoder to generate a SOT-style token sequence.
In other methods, SOT-style token sequences are used directly as CTC targets, forcing the encoder to produce speaker-specific features in SOT-style order~\cite{cse-net,sacrc}.
These approaches employ CTC to separate or reorder encoder outputs into SOT-style order, enabling the encoder to extract speaker-specific features and the decoder to align tokens to those outputs.

As discussed later in Section~\ref{enc-analysis}, our analysis of the behavior of SOT models reveals that while they accurately align tokens to their corresponding frames, they often misassign speakers.
This behavior suggests that the primary limitation of SOT lies in the speaker differentiation rather than the token alignment.
Based on this insight, we introduce Speaker-Distinguishable CTC (SD-CTC), a novel approach that enables an SOT model to distinguish speakers using only overlapping speech and each speaker's transcript---the same data used in standard SOT.
SD-CTC extends the CTC framework by incorporating a speaker discrimination task, assigning both a token and a speaker label to each frame.
Unlike existing methods that rely on separating or reordering encoder outputs, our approach focuses solely on speaker label prediction, similar to token prediction in standard CTC. This improves speaker attribution without requiring auxiliary information, which can be difficult to obtain in real-world scenarios.

Multi-task learning of SD-CTC and SOT helps the decoder generate a transcription for each speaker by leveraging encoder outputs that encode both frame-token and frame-speaker assignments.
SD-CTC introduces only a lightweight linear layer for speaker distinction, ensuring minimal impact on model size.
Moreover, the architecture remains unchanged even in many speaker scenarios, with only the output dimension of the linear layer changing to accommodate the maximum number of recognizable speakers. 
Since SOT effectively aligns tokens but struggles with speaker assignments, SD-CTC provides a more direct and effective approach to learning speaker distinctions without relying on auxiliary information.

Our contributions are twofold: i) we develop SD-CTC, which jointly assigns token and speaker labels, and ii) we demonstrate its competitive performance on the Libri\-SpeechMix dataset without auxiliary information.
These achievements contribute to improving the robustness of conversational speech recognition.


\section{Serialized Output Training Revisited}
\label{sec:sot}

\subsection{Original serialized output training}

SOT is a method for MT-ASR using an AED model, as illustrated on the left side of Figure~\ref{overview-fig}. 
For a set of utterances spoken by multiple speakers, 
the SOT model is trained to predict a token sequence that concatenates the transcriptions of all speakers in the order they begin speaking, with special tokens $\mathtt{{<}sc{>}}$ inserted to indicate speaker changes. 
Treating a multi-talker utterance as a single sequence eliminates the need to change the architecture as the number of speakers varies.
Additionally, using a single decoder to process utterances from multiple speakers improves recognition accuracy by leveraging contextual information from other speakers' outputs.

The SOT loss function is computed using cross-entropy, similar to single-speaker ASR.
Minimizing this loss allows the decoder's cross-attention mechanism to implicitly learn the association among acoustic frames, speakers, and tokens. 
It also optimizes the model parameters to ensure that the tokens are generated in the order in which speakers appear.

\subsection{Analysis of recognition errors in SOT models}
\label{enc-analysis}

Our analysis of the encoder outputs supports the hypothesis that the limitations of SOT models stem primarily from poor speaker differentiation rather than errors in token prediction.

\begin{figure}[tb]
  \begin{minipage}[b]{0.49\linewidth}
    \centering
    \includegraphics[keepaspectratio, scale=0.071]{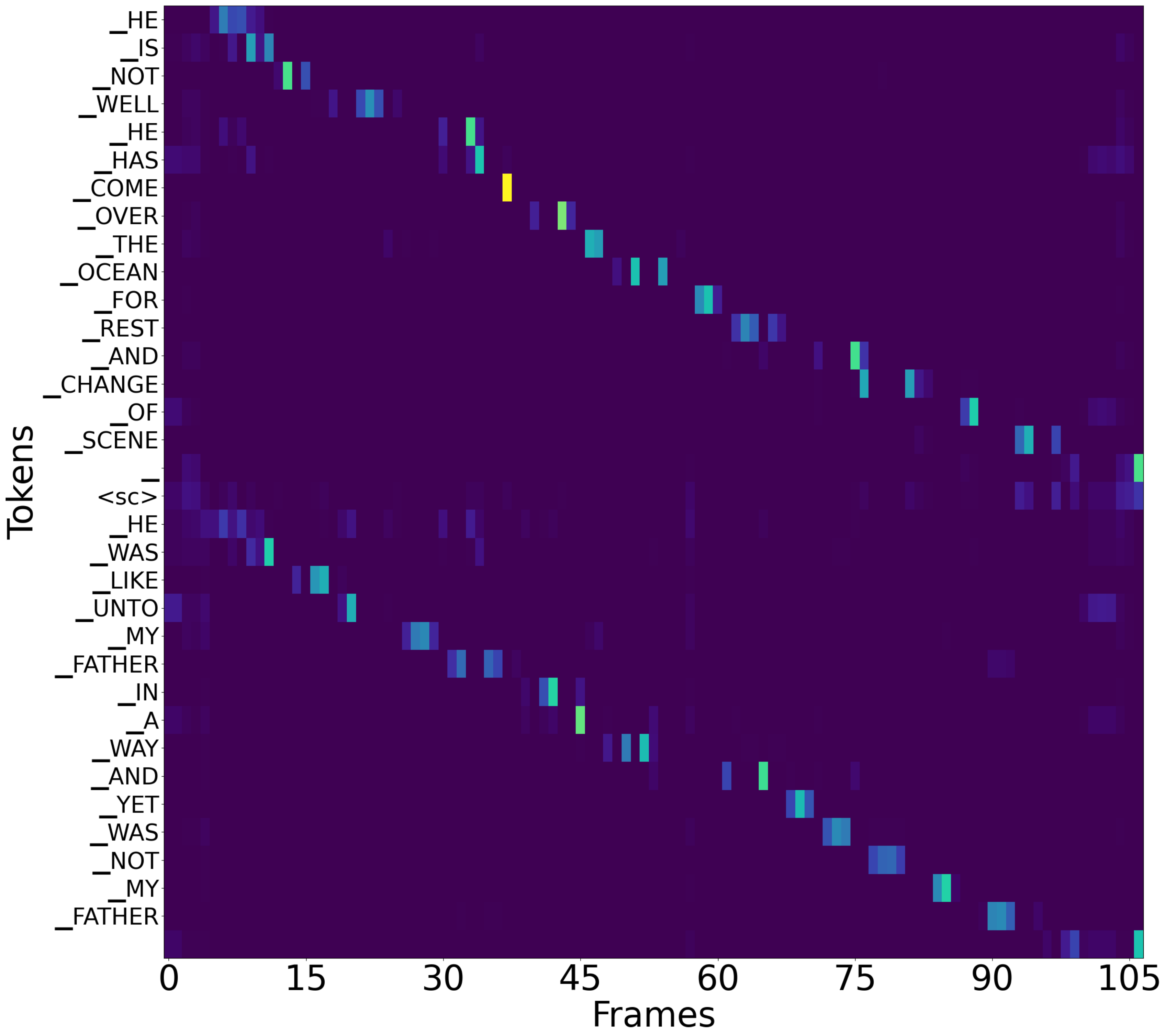}
    \subcaption{Last decoder attention map.}
    \label{last-layer-att}
  \end{minipage}
  \begin{minipage}[b]{0.49\linewidth}
    \centering
    \includegraphics[keepaspectratio, scale=0.115]{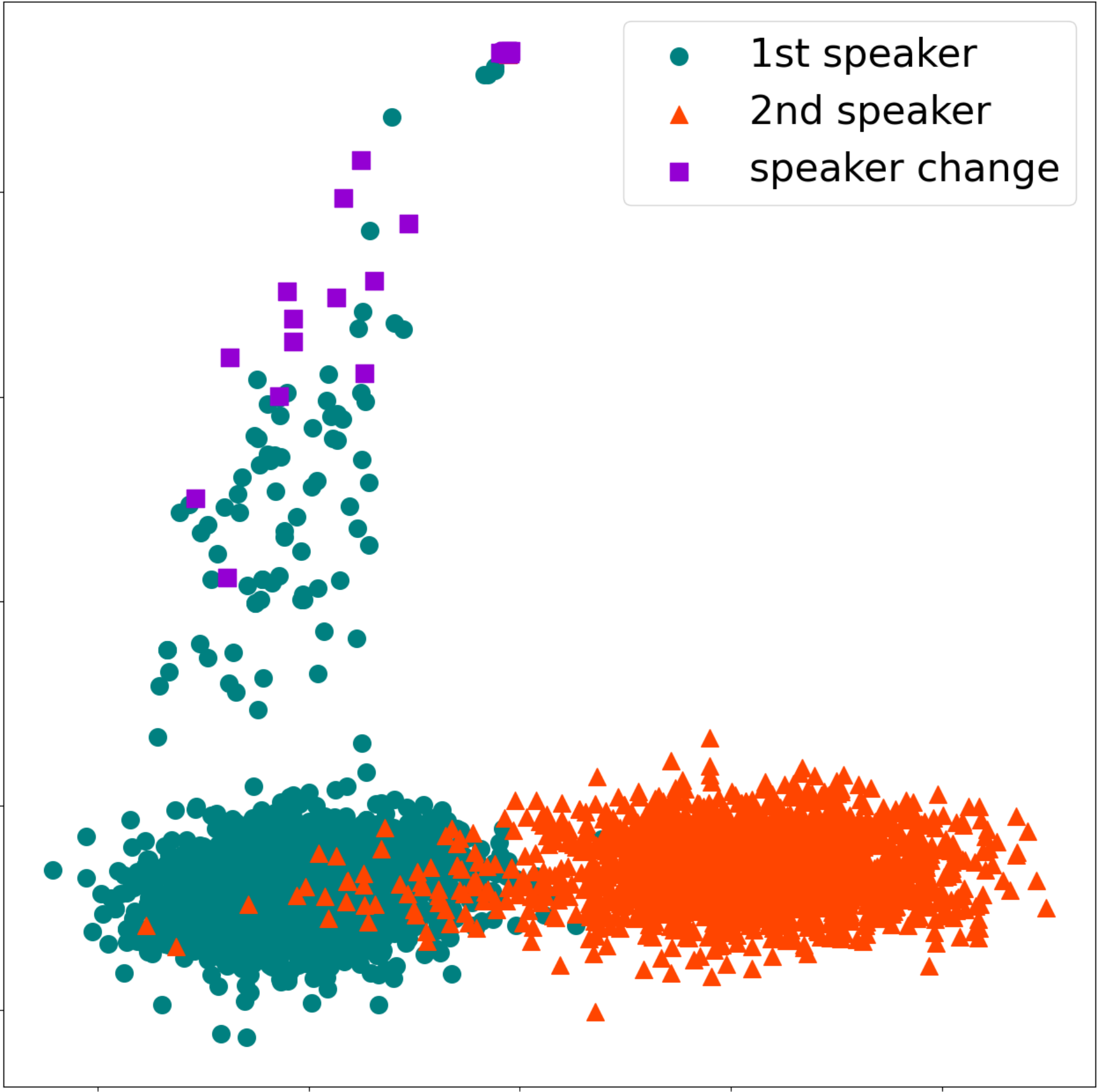}
    \subcaption{LDA of encoder outputs.}
    \label{lda-as-motivation}
  \end{minipage}
  \vspace{-.25cm}
  \caption{Analysis of recognition errors in SOT models. (a) Attention map of last decoder layer for test-clean-2mix-1541 in LibriSpeechMix. (b) LDA visualization of 100 samples from LibriSpeechMix, where green represents first speaker, orange represents second speaker, and purple represents $\mathtt{{<}sc{>}}$ token.}
  \vspace{-.5cm}
  \label{sot-anlysis}
\end{figure}

Figure~\ref{last-layer-att} presents an attention map of the final decoder layer, obtained via teacher-forcing on a sample where the SOT model omitted one speaker's transcription.
When provided with oracle tokens, the final layer correctly aligns acoustic frames with tokens, indicating that misalignment is not the source of the error.
Figure~\ref{lda-as-motivation} shows an LDA visualization of encoder outputs from 100 LibriSpeechMix samples, where the LDA class labels correspond to the speaker associated with the token in each frame.
The visualization reveals substantial overlap in the feature space, particularly between the representations of the first and second speakers, as well as between the first speaker and the $\mathtt{{<}sc{>}}$ token.
This overlap suggests that the model struggles to differentiate speakers, leading to errors such as misplacing tokens between speakers or miscounting speakers. 

%

\section{Proposed Method}

\subsection{Speaker-Distinguishable CTC}
\label{sd-ctc}

Speaker-Distinguishable CTC (SD-CTC) is a framework that enables the encoder to learn speaker-distinguishable representations using only overlapping speech and their corresponding transcriptions.
In contrast to prior CTC-based auxiliary losses for SOT~\cite{gencsep, sacrc}, SD-CTC avoids speaker extraction or frame reordering, focusing solely on learning speaker distinction.
Multi-task learning with the SOT loss (see Figure~\ref{overview-fig}) further aids the decoder in generating accurate transcriptions for each speaker.

In SD-CTC, the model estimates both a token and its associated speaker for each frame. 
Let $\bm{X} = (\bm{x}_1, \bm{x}_2, \dots, \bm{x}_T)$ be the acoustic feature sequence, where $\bm{x}_t$ is the feature vector at time $t$.
Let \(V\) denote the vocabulary, and let $\mathtt{{<}b{>}}$ represent the blank token for non-speech frames.
SD-CTC models the token probability distribution in the same way as the standard CTC as $ P_\mathrm{v}(\pi\mid\bm{x}_t) $,
where $ \pi \in  V\cup \{\mathtt{{<}b{>}} \} $. $P_\mathrm{v}(\mathtt{{<}b{>}}\, |\, \bm{x}_t)$  represents the probability of a non-speech frame.

SD-CTC extends this formulation by introducing a linear layer with a softmax function to the encoder for predicting the speaker probability distribution $P_\mathrm{s}(\sigma\mid\bm{x}_t)$, where $ \sigma \in S $. 
Let $S = \{\mathrm{s_1},\mathrm{s_2}, \dots, \mathrm{s}_M\} $ represents the speakers who are indexed in the order of their appearance within a speech segment containing overlapping utterances.

Unlike diarization, speaker-attributed ASR, or cross-utterance speaker classification, SD-CTC focuses solely on distinguishing speakers within a single speech segment.
It assumes that each frame belongs to only one speaker, similar to token assignments in standard CTC. 
This consistency aids the decoder in aligning speakers and tokens effectively.

\subsection{Speaker-specific token probability}

SD-CTC computes CTC loss separately for each speaker by introducing speaker-specific blank token ``$\mathtt{{<} \lnot s {>}}$'', which represents frames containing either non-speech or speech from a speaker other than the target speaker.
This token enables the model to ignore non-target speakers when computing the CTC loss. 
The extended token set is given by: $ V_{\textnormal{SD-CTC}} = V\cup \{\mathtt{{<} \lnot s {>}}\} $.
The speaker-specific token probability is then defined as
$ P(\sigma,\rho\mid\bm{x}_t) $,where $\sigma \in S$ and $\rho \in  V_{\textnormal{SD-CTC}}$, 
representing the probability that the speaker $\sigma$ emits the token ${\rho}$ at time $t$. 
This probability is computed as the joint probability of $P_\mathrm{s}(\sigma \, |\, \bm{x}_t)$ and $ P_\mathrm{v}(\pi \, |\, \bm{x}_t) $, obtained from the encoder's linear layers.
For non-blank tokens (i.e., $\rho \ne \mathtt{{<} \lnot s {>}}$), the speaker-specific token probability is given by
\begin{align}
     P(\sigma,\rho\mid\bm{x}_t)= P_\mathrm{s}(\sigma\mid\bm{x}_t) P_\mathrm{v}(\rho\mid\bm{x}_t). 
     \label{eq:rho_not_blank}
\end{align}
When $\rho = \mathtt{{<} \lnot s {>}}$, the probability of the speaker-specific blank token is computed as the sum of the non-speech probability $P_\mathrm{v}(\mathtt{{<}b{>}}\, |\, \bm{x}_t)$ and the probability of speech from any other speaker:
\begin{align}
  P(\sigma,  \mathtt{{<}\lnot s{>}}\mid\bm{x}_t)\hspace*{-7em}\notag\\
    &{} = P_\mathrm{v}(\mathtt{{<}b{>}}\mid\bm{x}_t) 
     + (1 - P_\mathrm{s}(\sigma\mid\bm{x}_t))
         (1 - P_\mathrm{v}(\mathtt{{<}b{>}}\mid\bm{x}_t)) \notag\\
    &{} = P_\mathrm{s}(\sigma\mid\bm{x}_t) \, P_\mathrm{v}(\mathtt{{<}b{>}}\mid\bm{x}_t)
    + (1 - P_\mathrm{s}(\sigma\mid\bm{x}_t)).
         \label{eq:rho_is_blank}
\end{align}
Thus, the final speaker-specific token probability is defined as follows:
\begin{align}
    \label{sd-ctc input}
    P(\sigma,\rho\mid\bm{x}_t)=
    \begin{dcases}
        P_\mathrm{s}(\sigma\mid\bm{x}_t) P_\mathrm{v}(\rho\mid\bm{x}_t) & (\rho \ne \mathtt{{<} \lnot s {>}} ), \\
        P_\mathrm{s}(\sigma\mid\bm{x}_t) P_\mathrm{v}(\mathtt{{<}b{>}}\mid\bm{x}_t) + (1 - P_\mathrm{s}(\sigma\mid\bm{x}_t)) \hspace*{-7.5em}\\
        & (\rho = \mathtt{{<} \lnot s {>}} ).
    \end{dcases}
\end{align}
In single-speaker scenarios, where only $\mathrm{s_1}$ is present, the model's predicted probability for that speaker $P_\mathrm{s}(\mathrm{s_1} \, |\, \bm{x}_t)$ is approximately one.
Thus, the right-hand side of Eq.~(\ref{sd-ctc input}) simplifies to the token prediction probability alone, which is equivalent to the standard CTC distribution.
This shows that SD-CTC naturally reduces to standard CTC in single-speaker scenarios.
\subsection{SD-CTC loss}

The CTC loss for each speaker is computed similarly to single-speaker ASR, using $P(\sigma,\rho\mid\bm{x}_t)$ as input.
For a speaker \(\sigma\) with the transcription $y_\sigma$, the CTC loss is defined as
\begin{equation}
L_{\textnormal{CTC}}(\sigma) = -\log P(\sigma, \bm{y}_\sigma \mid \bm{X}),
\end{equation}
where \(\bm{y}_\sigma\) is the correct token sequence for \(\sigma\).
The overall SD-CTC loss is then given by
\begin{equation}
L_{\textnormal{SD-CTC}} = 
\sum_{\sigma \in S}
L_{\textnormal{CTC}}(\sigma).
\end{equation}

\subsection{Techniques for learning and inference}
\label{tips-sd-ctc}
We adopt a two-stage schedule: multi-task pre-training with CTC/attention on single-speaker data, followed by fine-tuning on multi-talker data.
During pre-training, the speaker prediction linear layer is frozen with $P_\mathrm{s}(\mathrm{s_1} \, |\, \bm{x}_t) = 1$; during fine-tuning, the token prediction layer is frozen.
This approach reduces training time for multi-talker data and improves training stability.
With these parameters fixed, the token prediction probability $P_\mathrm{v}(\pi|\bm{x}_t)$ serves as a weight for each speaker at each frame. 
The model is then trained to predict speakers who have a high probability of producing their own tokens, thereby learning to distinguish among different speakers.
During inference, we perform beam search using only the attention decoder to generate SOT-style hypotheses.
For each of the top $K$ hypotheses, we compute the SD-CTC log-likelihood for the corresponding speaker transcript and add it to the decoder’s log-likelihood for re-scoring, improving accuracy.

\begin{table*}[t]
\centering
\caption{
MT-ASR Performance (cpWER \%) on Libri\-Speech and Libri\-SpeechMix evaluation sets. 
Column ``Auxiliary info.'' indicates any additional information required for training/decoding.
In each set, the best score is shown in bold, and the second-best score is underlined.
Results for methods reproduced in this study appear below broken line.
}
\setlength{\tabcolsep}{2pt}
\vspace{-0.3cm}
\label{tab:comp}
\resizebox{.75\linewidth}{!}{%
\begin{tabular}{@{}lccccc@{}}
\bhline{1pt}
\multirow{2}{*}{\textbf{Method}} & \begin{tabular}[c]{@{}c@{}} \textbf{\# of} \\ \textbf{params.} \end{tabular} & \begin{tabular}[c]{@{}c@{}} \textbf{Auxiliary} \\ \textbf{info.}\end{tabular} & \begin{tabular}[c]{@{}c@{}} \textbf{LibriSpeech} \\ \textbf{test-clean} \end{tabular} &  {\begin{tabular}[c]{@{}c@{}} \textbf{LibriSpeechMix} \\ \textbf{2mix} \end{tabular}} \\ \hline
t-SOT~\cite{t-sot} & 139M & token-level timestamp & 3.3 & 4.4 \\
SA-SOT~\cite{sa-sot}  & 136M & token-level timestamp & \underline{2.5}  & \textbf{3.4}  \\
MT-RNNT-AFT~\cite{mt-rnnt-aft}  & 120M & not necessary & 2.6  & 3.7  \\ \hdashline
SOT Baseline~\cite{sot}  & 114M & not necessary & 3.0  & 4.7  \\
SOT + SACTC~\cite{sacrc} & 114M & not necessary & 2.7  & 5.4  \\
SOT + GEncSep~\cite{gencsep}  & 125M & not necessary & 2.6  & 3.9  \\
SOT + SD-CTC (proposed) & 114M & not necessary & \textbf{2.4}  & \underline{3.5}  \\
\quad  $\hookrightarrow$ AED only inference (as ablation study) &  &  & \textbf{2.4}  & 4.1  \\
\quad  $\hookrightarrow$ CTC only inference (as ablation study) &  &  & 6.6  & 9.5  \\
\bhline{1pt}
\end{tabular}%
}%
\vspace{-0.5cm}
\end{table*}

\section{Experiments}
We evaluate our method using ESPnet~\cite{espnet}, training on Libri\-Speech~\cite{librispeech} and testing on Libri\-SpeechMix~\cite{sot}.


\subsection{Experimental setups}

The training procedure followed the approach in~\cite{t-sot}.
Specifically, for each Libri\-Speech sample, a second sample was randomly selected and mixed with a randomly determined delay. 
The number of mixed speech samples was also chosen randomly, with a maximum of two. 
Since the samples were selected and mixed on the fly, the mixing combinations varied across epochs.

Data augmentation techniques included speed perturbation~\cite{speed-perturbation} (0.9, 1.0, 1.1), volume perturbation~\cite{volume-perturbation} with (0.125x to 2.0x), and Adaptive SpecAugment~\cite{specaugment} with the Libri\-Full\-Adapt policy~\cite{adaptive_specaugment}.
The evaluation was conducted on Libri\-SpeechMix (containing two-speaker mixtures) and the Libri\-Speech  test-clean set (containing single-speaker recordings).

The SOT model architecture comprised a 12-layer Conformer encoder~\cite{conformer} and a 6-layer Transformer decoder, with each layer having a 512-dimensional hidden size, eight attention heads, and a 2048-dimensional fully connected layer. 
In the proposed method, additional linear layers were introduced for token prediction and speaker distinction.
For comparison, we employed GEncSep~\cite{gencsep} and SACTC~\cite{sacrc}, which are auxiliary CTC-based loss functions for SOT. 
In our GEncSep implementation, the separator used a 512-dimensional, two-layer BiLSTM.
Unlike~\cite{gencsep}, where a fixed number of speakers was assumed, our implementation allowed for an arbitrary number of speakers by treating all frames as blank for the second speaker when processing single-speaker speech. 

The training process comprised two stages: CTC/Attention pre-training on single-speaker data, followed by training on both single- and two-speaker data.
In the second stage, CTC was not used in the baseline SOT model, whereas the other methods employed multi-task CTC/Attention learning. 
The decoder used 5000 subwords, whereas CTC branches used 100 subwords to enhance accuracy~\cite{char-ctc}.

Mini-batches were constructed using ESPnet's numel scheme, with batch bins set to 180000000 (resulting in an average batch size of 259).
The AdamW optimizer was used with an initial learning rate of 0.0005 and a Cosine Annealing scheduler with a warmup step of 2048.
Training was conducted for 60 epochs in the first stage and 216 epochs in the second stage.
The evaluation model was obtained by averaging the ten models with the highest validation accuracy. 
The weight of each CTC loss variant was set to 0.3. 
During inference, beam search was performed with a beam width of 16, and hypotheses were re-scored using the CTC score with a weight of 0.3 for all methods except the baseline SOT.
Language models were not used to isolate the impact of the auxiliary task in the SOT model. 

For evaluation, we employed the concatenated minimum-permutation WER (cpWER)~\cite{cpwer}.
This metric concatenates the output texts for all speakers and compares them with concatenated reference texts, selecting the lowest WER among all possible permutations.
Since SD-CTC cannot assign multiple speakers to a frame or track speakers across segments, DER is not an appropriate metric for evaluating speaker discrimination. 
Instead, cpWER was used to simultaneously assess both speaker discrimination and token prediction accuracy.

\subsection{Evaluation results}

The experimental results are summarized in Table~\ref{tab:comp}.
Results for methods utilizing separated decoders~\cite{mt-rnnt-aft} and auxiliary information~\cite{t-sot,sa-sot} are shown above the broken line, whereas the results for methods tested in this study appear below.
The table shows accuracy for AED-only inference and CTC-only inference in the model trained with the proposed method.

A comparison between SOT Baseline and SOT + SD-CTC (AED-only inference) shows that SD-CTC enhances the SOT decoder performance, suggesting that improved speaker distinction in the encoder reduces speaker-related errors.
Moreover, with SD-CTC re-scoring during inference, the proposed method achieved a 26\% relative reduction in error rate compared with SOT, attaining a cpWER of 3.5\% for two-speaker speech. 

These findings demonstrate that SD-CTC achieves comparable performance with fewer parameters than SA-SOT, a state-of-the-art method relying on auxiliary information. 
A comparison with SACTC and GEncSep suggests that SD-CTC is superior in performance and more parameter-efficient than methods based on acoustic-information separating or reordering.

\subsection{Analysis}

\begin{figure}[tb]
  \begin{minipage}[b]{0.49\linewidth}
    \centering
    \includegraphics[keepaspectratio, scale=0.123]{base_encoder_out_lda_100_superbig.pdf}
    \subcaption{w/o SD-CTC}
    \label{lda-a}
  \end{minipage}
  \begin{minipage}[b]{0.49\linewidth}
    \centering
    \includegraphics[keepaspectratio, scale=0.123]{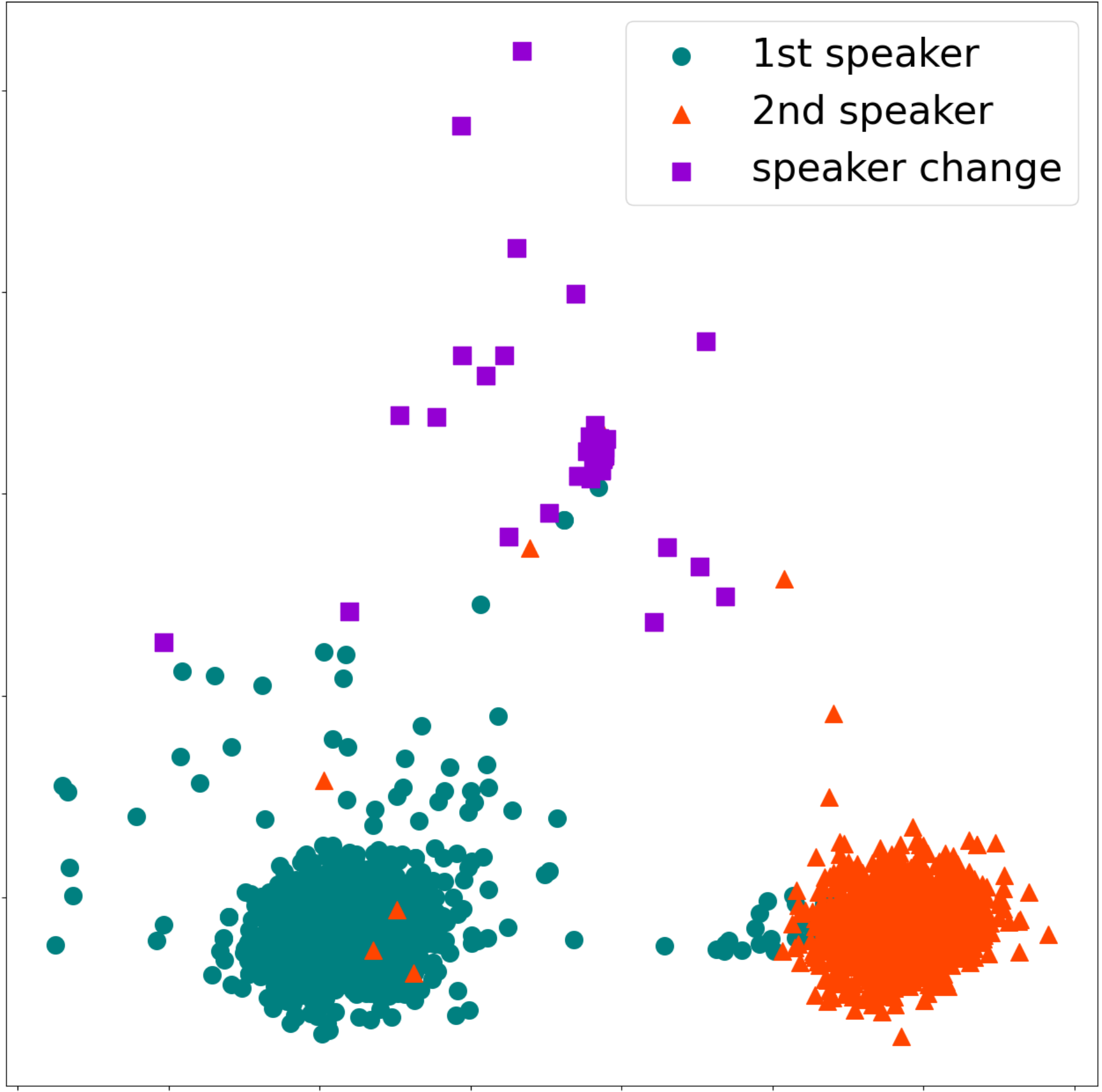}
    \subcaption{w/ SD-CTC}
    \label{lda-b}
  \end{minipage}
  \vspace*{-.25cm}
  \caption{LDA visualization of encoder outputs attended by each token where green represents first speaker, orange represents second speaker, and purple represents $\mathtt{{<}sc{>}}$ token.}
  \vspace{-.25cm}
  \label{lda}
\end{figure}

To assess the model's ability to distinguish between speakers, we visualized the encoder output vectors using the method described in Section~\ref{enc-analysis}.
As shown in Figure~\ref{lda}, the proposed method produces more widely separated clusters, each corresponding to a specific speaker or to the $\texttt{{<}sc{>}}$ token, indicating that the encoder more effectively differentiates speakers.
 
\begin{figure}[tb]
  \begin{minipage}[b]{0.465\linewidth}
    \centering
    \includegraphics[keepaspectratio, scale=0.0615]{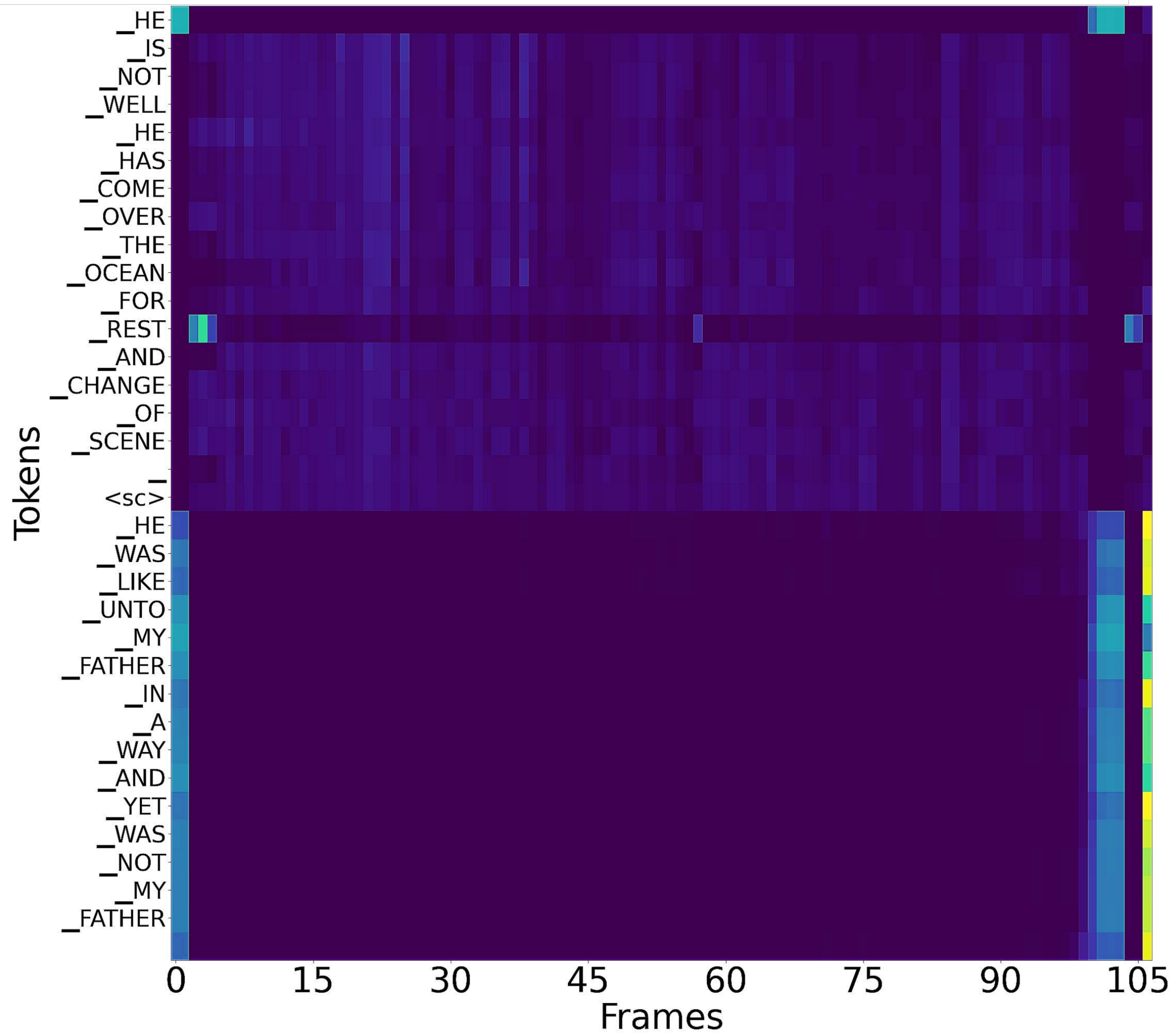}
    \subcaption{w/o SD-CTC}
    \label{2nd-att-comp-a}
  \end{minipage}
  \begin{minipage}[b]{0.525\linewidth}
    \centering
    \includegraphics[keepaspectratio, scale=0.061]{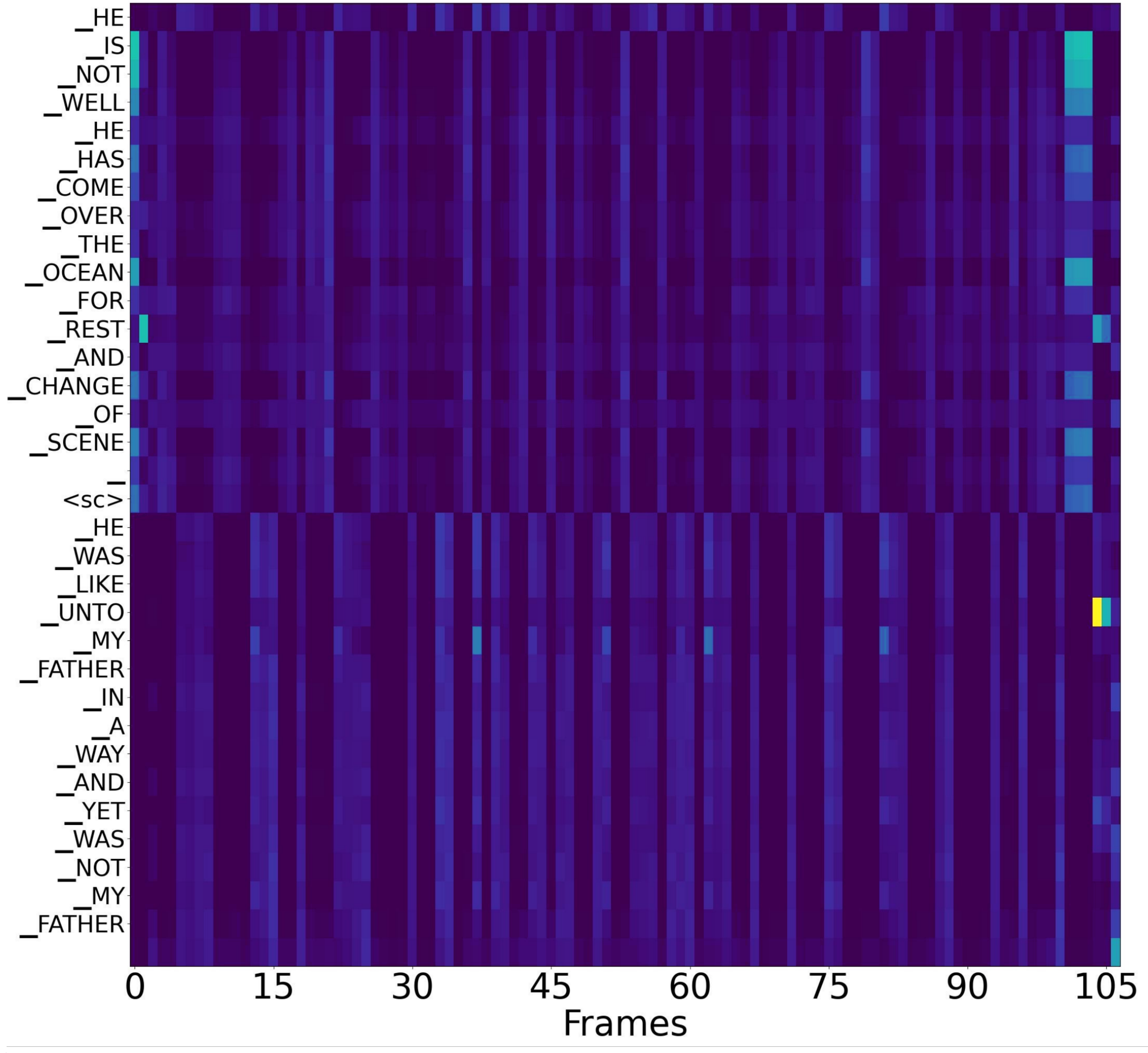}
    \subcaption{w/ SD-CTC}
    \label{2nd-att-comp-b}
  \end{minipage}
  \vspace{-.6cm}
  \caption{Attention map of second decoder layers. Target utterance is test-clean-2mix-1541 in LibriSpeechMix.}
  \vspace*{-.5cm}
  \label{2nd-att-comp}
\end{figure}

We observed that the frame–speaker assignments were clearly reflected in the cross-attention patterns of the low-level decoder. Figure~\ref{2nd-att-comp} presents the attention maps of the second decoder layer on the same sample used in Figure~\ref{enc-analysis}.
Notably, whereas the SOT model attends to one speaker's tokens, the model trained with the proposed method attends separately to each speaker's tokens, effectively assigning frames to the correct speaker.
These visualizations suggest that the encoder has learned to distinguish speakers at a frame-level and that the decoder leverages this information to enhance speaker separation.

\section{Conclusions}
In this paper, we propose Speaker-Distinguishable CTC, which improves multi-talker ASR accuracy  using only overlapping speech and each speaker’s transcript. Our evaluation experiments demonstrate that multi-task learning with SOT and SD-CTC significantly improves recognition accuracy, achieving performance equivalent to state-of-the-art methods that rely on auxiliary information. These results suggest that large-scale real conversation corpora --- where auxiliary information is difficult to obtain --- can be effectively used to further improve overlapping speech recognition accuracy.

\bibliographystyle{IEEEtran}
\bibliography{mybib}

\end{document}